\documentclass[prl,aps,floats,showpacs]{revtex4}

\usepackage{graphicx}

\newcommand{\be}{\begin{equation}}
\newcommand{\ee}{\end{equation}}  
\newcommand{\bea}{\begin{eqnarray}}
\newcommand{\eea}{\end{eqnarray}}

\newcommand{\rr}{{\bf r}}

\begin{document}

\title{ Monte Carlo simulation method
        for Laughlin-like states in a disk geometry}

\author{ Orion Ciftja and Carlos Wexler }

%\address{
\affiliation{
         Department of Physics and Astronomy, University of
         Missouri-Columbia, Columbia, Missouri 65211}

\begin{abstract}
We discuss an alternative accurate Monte Carlo method to calculate the 
groundstate energy and related quantities for Laughlin states of the
fractional quantum  Hall effect in a disk geometry.
This alternative approach allows us to obtain accurate
bulk regime (thermodynamic limit) values for various quantities
from Monte Carlo simulations with a small number of particles (much
smaller than that needed with standard Monte Carlo approaches).
\end{abstract}

%\date{\today}   

\pacs{73.43.-f, 05.30.Fk, 71.70.Di}

\maketitle

\section{Introduction}

The discovery of the fractional quantum Hall effect (FQHE) has
stimulated extensive studies on the properties of two-dimensional (2D)
quantum many-electron systems in a strong magnetic field \cite{tsui}.
It is now understood that the FQHE represents the condensation of
nearly 2D electrons subject to a strong perpendicular magnetic
field (at low enough temperatures and low enough amount of disorder)
into an incompressible quantum fluid formed at some
specific filling factors.
A neutralizing positive charge density is present to
preserve overall charge neutrality and, to lowest approximation, can be
thought of as a uniform positive density in the same plane as the 2D
electrons (in reality the positive charges are the ionized donors that
are roughly distributed randomly a {\em spacer thickness} away
and produce a small amount of disorder and an overall constant
shift in the energies).

The strong magnetic field quantizes the electrons's motion on the plane and
quenches the kinetic energy of each electron to a discrete set of
Landau levels (LL) separated by the relatively large cyclotron energy 
$\hbar \omega_c = \hbar e B/m^*$, where $-e(e>0)$ is the electron charge and
$m^*$ is the effective mass of electrons in the semiconductor ($m^*
\simeq 0.07 m_e$ in GaAs, where $m_e$ is the bare electron's mass). 
In addition, the Zeeman splitting
spin-polarizes the electrons rendering them {\em effectively
spinless}.  
In each LL, $\rho_{LL} = 1/2\pi l_0^2$ ($l_0 = \sqrt{\hbar/e B}$ 
is the {\em magnetic length}) electrons per unit area can be
accomodated.  It is evident that for large enough magnetic fields only 
the lowest LL (LLL) is occupied (and only the lowest spin sub-band),
and if the occupation is not complete the system is highly degenerate.
At particular {\em filling factors} $\nu = \rho/\rho_{LL}$,
electron interactions lead to highly correlated states which exhibit
an excitation gap and result in the observed FQHE.  These particular
filling factors form a hierarchy, of which the simplest FQHE states
have filling factors $\nu = 1/m$ with $m=3$ and $5$.

For filling factors of the form $\nu=1/m$ (m odd) the 
unnormalized Laughlin \cite{laughlin} trial wave function for $N$
electrons can be written as:
\begin{equation}
\Psi_{m}(z_1 \ldots z_N)=\prod_{i<j}^{N} (z_i-z_j)^m \ \prod_{i=1}^{N}
e^{-\frac{|z_i|^2}{4 l_0^2}} \ ,
\label{laughlin}
\end{equation}
where $z_{j}=x_{j}+i y_{j}$ is the position of the $j$-th electron in complex
coordinates. This wave function gives an excellent description of the
true ground state of the electrons for $m=3$ and $5$.
For $m \geq 7$ the electrons tend to form a Wigner crystal~\cite{lam}
consistent with the experimental observation~\cite{mendez} 
that the FQHE does not occur for filling factors $\nu \leq 1/7$.

Since the Laughlin wave function
lies entirely in the LLL, the expectation value of the kinetic energy
per electron 
\begin{equation}
\frac{\langle \hat{K} \rangle}{N}=\frac{1}{2} \hbar \omega_c \ ,
\label{kinetic}
\end{equation}
is constant and becomes irrelevant, therefore the only important
contribution of the quantum mechanical Hamiltonian
$\hat{H}=\hat{K}+\hat{V}$ originates from the total potential energy
operator: 
\begin{equation}
% \hat{H}=\hat{K}+\hat{V} \ \ \ ; \ \ \
\hat{V}=\hat{V}_{ee}+\hat{V}_{eb}+\hat{V}_{bb} \ , 
\label{pot_en}
\end{equation}
where
\be
\hat{V}_{ee}=\sum_{i<j}^{N} \frac{e^2}{|\rr_i-\rr_j|} \ \ \ ; \ \ \
\hat{V}_{eb}=-\rho_0 \sum_{i=1}^{N} \int_{\Omega_N} d^2r \ 
\frac{e^2}{|\rr_i-\rr |} \ \ \ ; \ \ \
\hat{V}_{bb}=\frac{\rho_0^2}{2} \int_{\Omega_N} d^2r \int_{\Omega_N} 
d^2r^{\prime} \
\frac{e^2}{|\rr-\rr^{\ \prime}|} \ ,
\label{potential}
\ee
are the electron-electron, electron-background and background-background
potential energy operators respectively. Here we have assumed a simple
geometry approprite for the circular symmetry of the Laughlin
wave function, where a positive background density 
$\rho_0=\nu/(2 \pi l_0^2)$ is spread over a disk $\Omega_N$ of radius
$R_N = l_0 \sqrt{2 N/\nu}$ (i.e. it cancels the electronic density in 
the thermodynamic limit and makes the system neutral for all $N$).

%%%%%%% techniques %%%%%%%%%

Numerous techniques have been employed to calculate the expectation
value of the potential energy per particle $\langle \hat{V} \rangle/N$
[see Eq.\ (\ref{pot_en})] in the Laughlin state [Eq.\
(\ref{laughlin})]. For example, Laughlin \cite{laughlin} initially
employed the hypernetted-chain method to estimate the value of this
correlation energy with a $\sim 1$\% accuracy); and various standard 
Monte Carlo (MC) schemes have been proposed, all of which are essentially
exact in the thermodynamic limit.
 
An excellent description of a standard MC computation 
of the potential energy and other relevant quantities
in a disk geometry
is given by Morf and Halperin \cite{morf}.  Spherical geometries are
also used quite often since the convergence to the thermodynamic limit
is quicker because boundary effects are eliminated \cite{spherical}.     

Although considerable more care is needed in the disk geometry to
eliminate boundary effects when extrapolating (necessarily) finite-$N$
results to the thermodynamic limit (in particular due to the
long-range nature of the Coulomb potential), there are cases in which
the spherical geometry is either inconvenient, or plainly incompatible
with the state under consideration (for example for the study of
possible quantum Hall nematic phases \cite{brs} for which considerable
topological defects would be generated at the poles of the sphere).

Furthermore, the value of the correlation energy in the thermodynamic limit is not
easily extracted from standard MC simulation (see Sec.\
\ref{sec:standard}) data, since the limit is approached very slowly,
with corrections of the  order of $1/\sqrt{N}$, requiring repeated
calculations for various $N$ and a careful extrapolation of the
results to the $N \rightarrow \infty$ limit. It is, therefore, highly
desirable to explore methods that would expedite the extrapolation to
the thermodynamic limit.  

In Sec.\ \ref{sec:standard} we describe, for the sake of completeness,
the procedure used in the standard MC approach.  Section \ref{sec:center}
describes an alternative method that converges to the thermodynamic
limit considerably faster.  We discuss our results in Sec.\
\ref{sec:discussion}.

%%%%%%%%%%%%%%%%%%%%%%%%%%%%%%%%%%%%%%%%%%%%%%%%%%%%%%%%%%%%%%%%%%%%%
\section{The standard Monte Carlo approach}
\label{sec:standard}

In the standard MC approach one considers the calculation of the
expectation value of the potential energy operators as given in Eq.\
(\ref{potential}).
The background-background interaction potential poses no problem, it
can be calculated analytically and is found to be
\begin{equation}
\frac{\langle \hat{V}_{bb} \rangle}{N}=\frac{1}{N}
\frac{\rho_0^2}{2} \int_{\Omega_N} d^2r \int_{\Omega_N} d^2r^{\prime} \
\frac{e^2}{|\vec{r}-\vec{r}^{\ \prime}|}=
\frac{8}{3 \pi} \sqrt{\frac{\nu N}{2}} \frac{e^2}{l_0}  \ .
%\frac{8}{3 \pi} \sqrt{\frac{N}{2m}} \frac{e^2}{l_0} \ \ \ ; \ \ \
%\nu=\frac{1}{m} \ .
\label{bbintegral}
\end{equation}

%%%%%%%%%%%%%%%%%%%%%%%%%%%%%%%%% eb %%%%%%%%%%%%%%%%%%%%%%%%%%%%%   

\begin{figure}[!htb]
\includegraphics{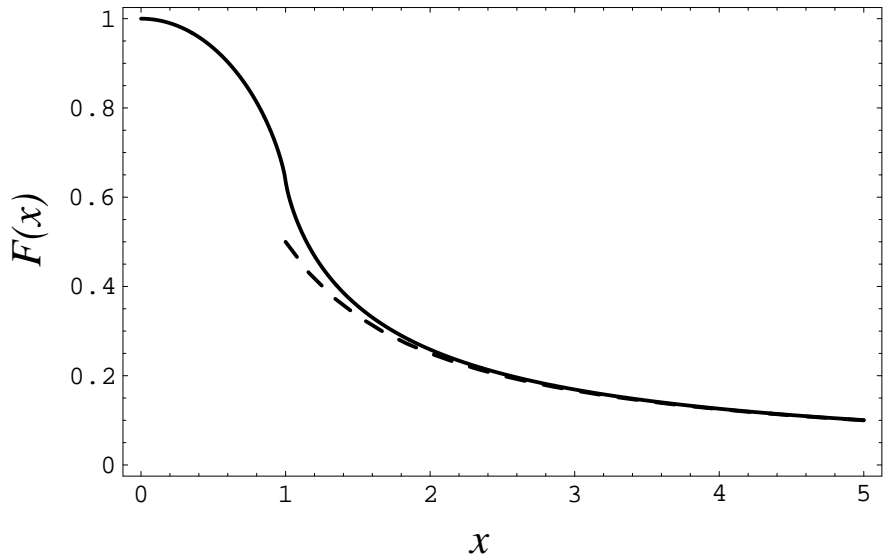}
\caption[]{The electron-background function $F(x)$.  Here
	   $x=r/R_N$, where $r$ is the distance of an
           electron from the center of the disk and
           $R_N$ is the radius of the finite disk filled with
           neutralizing
	   positive background. The dotted line
	   shows the $1/(2x)$ asymptotic dependence.  Most electrons 
	   sit in the $x \leq 1$ region.}
\label{figeb}
\end{figure}

In order to compute the expectation value of the electron-background
interaction potential one conveniently writes it as 
\begin{equation}
\hat{V}_{eb}=\sum_{i=1}^{N} \hat{v}_{eb}(\rr_i) \ \ \ ; \ \ \
      \hat{v}_{eb}(\rr_i)=-\rho_0 \ \int_{\Omega_{N}}\!\! d^2r \
       \frac{e^2}{|\rr_i-\rr|} \ ,
\label{veb_onebody}
\end{equation}
where $\hat{v}_{eb}(\rr_i)$ is the interaction potential of
a single electron at position $\vec{r}_i$  with the
uniform positive background in the finite disk. 
Such electron-background interaction potential depends on the ratio
$r_i/R_N$, where $r_i=|\rr_i|$ is electron's distance from
the center of the disk and $R_N$ is the radius of the finite disk
and can be expressed as
\begin{equation}
   \hat{v}_{eb}(\rr_i)=
            -\sqrt{2 \nu N} \ F(r_i/{R_N}) \
            \frac{e^2}{l_0} 
  \ \ ; \ \ 
F(x)=\int_{0}^{\infty} dy \ \frac{J_{0}(x \ y) J_{1}(y)}{y} = 
\left\{
\begin{array}{l}
\displaystyle \frac{2 E(x^2)}{\pi} \ , \ \ x \le 1  \\ \\
\displaystyle \frac{_2F_1(\frac{1}{2},\frac{1}{2};2;\frac{1}{x^2})}{2x}
\ , \ \ x \ge 1
\end{array}
\right. \, ,
\label{feb_onebody}
\end{equation}
where $J_n(x)$ are Bessel functions of order $n$, $E(x)$ is the
complete elliptic integral, and $_2F_1(a,b;c;z)$ is the
hypergeometric function.  Figure~\ref{figeb} plots the function
$F(x)$.  It is interesting to note that $F(0)=1$, $F(1)=2/\pi$ and 
$F(x)\sim 1/(2x)$ for $x \gg 1$, as expected for the Coulomb potential 
far from a charged disk.  Although $F(x)$ can be expressed
analytically, it generally preferable to store it in a table, and
interpolate it for all $x$ for all the calculations that follow. 
                                                                           
While most electrons stay within the confines of the neutralizing
background (i.e.\ $x \le 1$), electrons near the edge may spread outside
the disk to some extent (although it is extremely unlikely that they will
spread to more than a few magnetic lengths from the edge).
The expectation value of the electron-background interaction
potential during the MC simulation can then be calculated using         
\begin{equation}
\frac{\langle \hat{V}_{eb} \rangle}{N}=\frac{1}{N} \
\left\langle \sum_{i=1}^{N} \hat{v}_{eb}(\rr_i) 
\right\rangle   \ .
\label{ebpot}
\end{equation}
Finally, the expectation value of $\hat{V}_{ee}$ is accordingly given by
\begin{equation}
\frac{\langle \hat{V}_{ee} \rangle}{N}=\frac{1}{N}
\left\langle 
\sum_{i<j}^{N} \frac{e^2}{|\rr_i-\rr_j|}
\right\rangle \ .
\label{eepoti}
\end{equation}

In the usual Metropolis MC method \cite{metropolis}, the expectation
value of an operator can be computed by averaging the value of the
operator over numerous configurations 
$\{\vec{r}_1, \ldots ,\vec{r}_N \}$ of the
many-body system that obey detailed balance, that is, the probability
ratios between pairs of discrete configurations are related by the
ratios of {\em the} probability distribution for the system [in this
case $|\Psi_m(z_1,\ldots,z_N)|^2$, see Eq.\ (\ref{laughlin})].
Usually several million configurations are used for each $N$ and the
results are extrapolated to the thermodynamic limit by considering a
sequence of various increasing $N$-s.
 
A MC step (MCS) consists of attempts to move one by one all the
electrons of the system by a small distance of order $\Delta$ in a
random direction.  After each attempt (to move the $i$-th electron
from $\vec{r}_{i}^{\ old}$ to $\vec{r}_{i}^{\ new}$,
the probability ratio between the ``new'' state and 
the ``old'' state is then computed:
\begin{equation}
\frac{|\Psi(\rr_1, \ldots \rr_{i}^{\rm new} \ldots \rr_{N})|^2}{
      |\Psi(\rr_1, \ldots \rr_{i}^{\rm old} \ldots \rr_{N})|^2}=
\exp{\left[ m \sum_{j \neq i}^{N} \left(\ln |\rr_j-\rr_{i}^{\rm new}|^2 
                               -\ln |\rr_j-\rr_{i}^{\rm old}|^2 
      \right) \right]} \cdot
\exp{\left[-\frac{1}{2 l_0^2} 
   \left(|\rr_{i}^{\rm new}|^2-|\rr_{i}^{\rm old}|^2 \right) \right]} \ .
%\exp{\left[-
%   \left(\frac{|\rr_{i}^{new}|^2}{2 l_0^2}-
%          \frac{|\rr_{i}^{old}|^2}{2 l_0^2} \right) \right]} \ 
\end{equation}
In the usual Metropolis scheme \cite{metropolis}, if this ratio is
bigger than a uniformly distributed number in the [0,1] range the
attempt is accepted, otherwise it is rejected.  The parameter $\Delta$
is adjusted so that the acceptance ratio is close to $50 \%$.
After attempting to move all electrons (one MCS),
the electron configurations are then used to calculate the operator
under consideration.  Averaging over numerous MCS-s converges
gradually (as $1/\sqrt{{\rm number \ of \ MCS}}$) to the desired
expectation value.
Normally it is convenient to disregard numerous (several thousand)
initial configurations to reach a good  ``thermalization'' before the
averaging begins, which significantly reduces the expurious effects of
the somewhat arbitrary initial configurations.                               
All the results that we report here were obtained after discarding 100,000 
``thermalization'' MCS-s and using 
$2 \times 10^6$ MCS-s for averaging purposes.                      

In Table \ref{tabstandard} we show the correlation energy per particle 
for finite systems of $N$ electrons in the Laughlin states $m=3$ and
$m=5$ obtained using the standard MC method described above.
The results are rounded in the last digit.

\begin{table}
\caption[]{Correlation energy per particle in the Laughlin state
           for filling factors $\nu=1/3$ and $1/5$.
           These results were obtained after a standard Monte Carlo
           simulation in a disk geometry.
           Energies are in units of $e^2/l_0$.}
\label{tabstandard}
\begin{center}
\begin{tabular}{|c|c|c|}
\hline                                                  
N         & m=3                 & m=5              \\ \hline
4         & -0.38884             & -0.32159            \\ \hline
16        & -0.39766             & -0.32328            \\ \hline
36        & -0.40129             & -0.32446            \\ \hline
64        & -0.40323             & -0.32510           \\ \hline
100       & -0.40445             & -0.32550            \\ \hline
144       & -0.40521             & -0.32577            \\ \hline
196       & -0.40579             & -0.32594            \\ \hline
400       & -0.40675             & -0.32624            \\ \hline
\end{tabular}
\end{center}
\end{table}

\begin{figure}[!htb]
\includegraphics{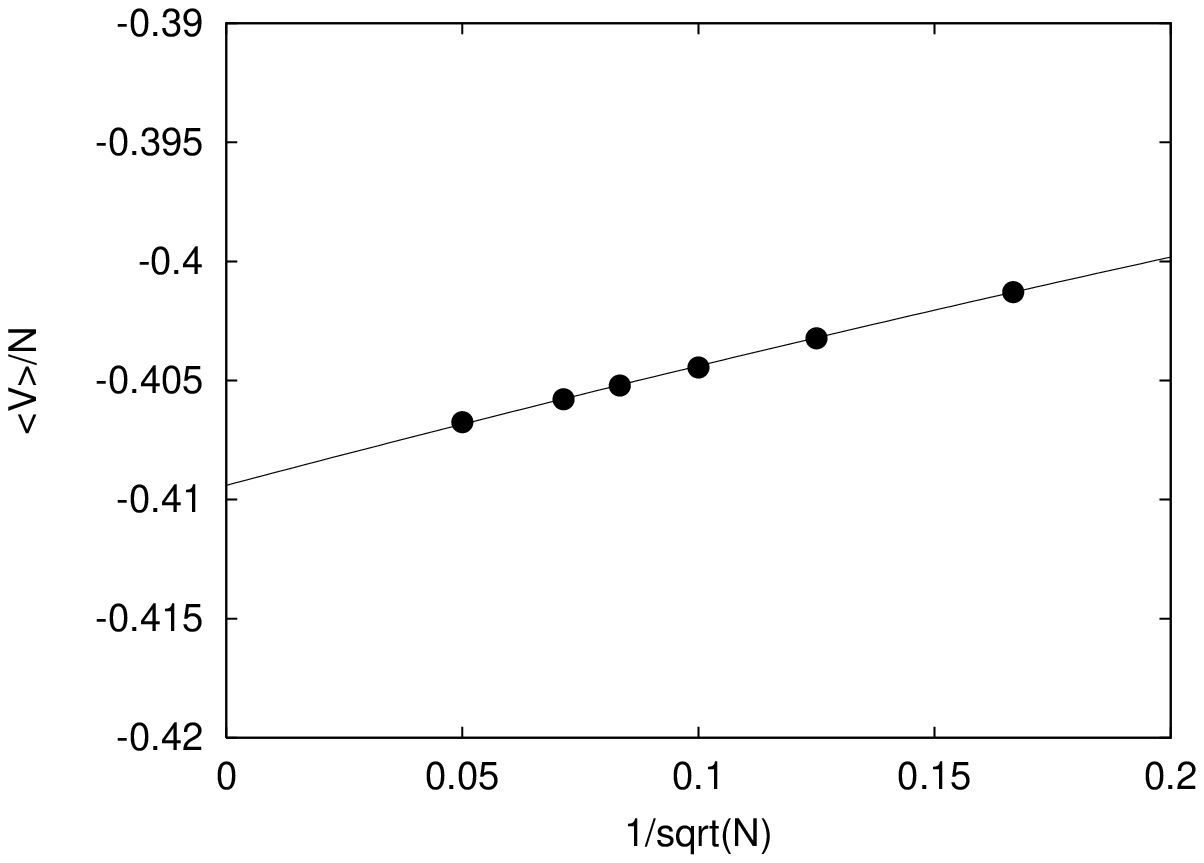}
\includegraphics{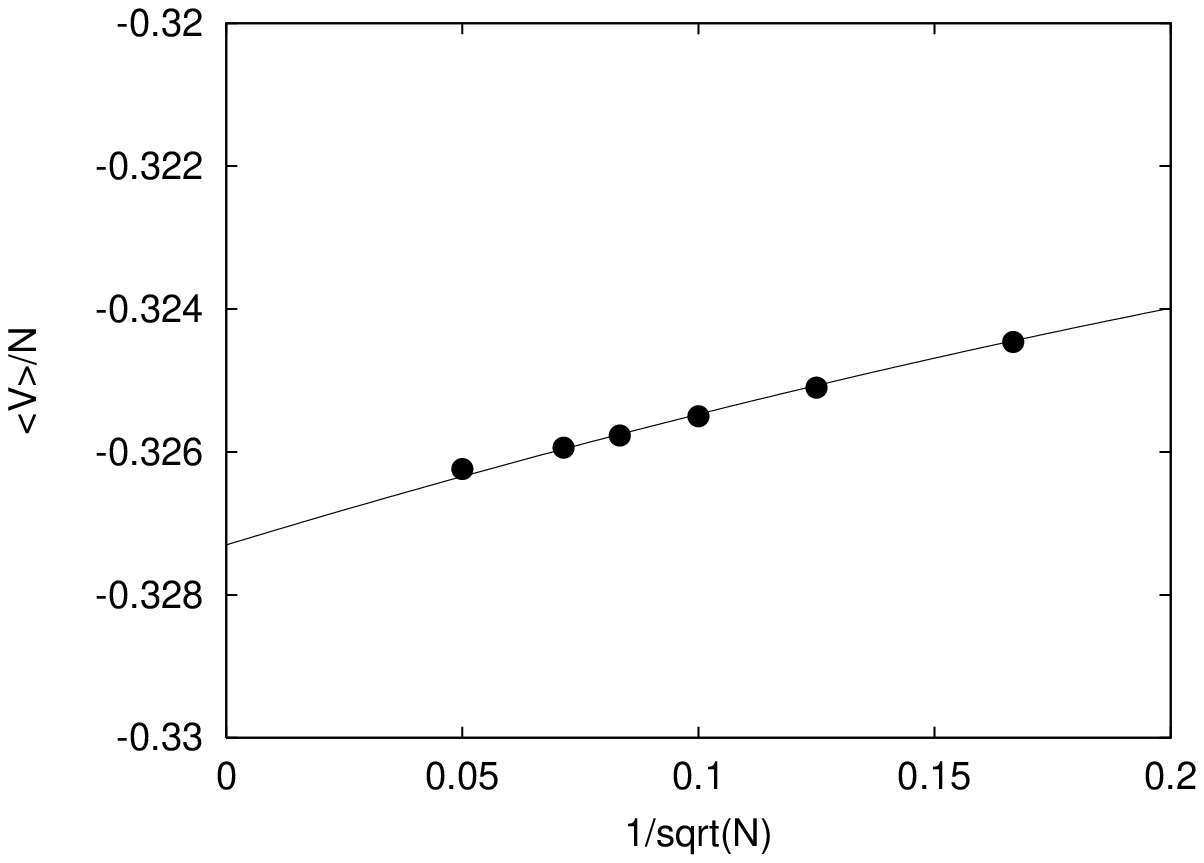}
\caption[]{ Monte Carlo results in disk geometry for the
            Laughlin state at 
            %   $\nu=1/3$. 
            $\nu=1/3$ (top panel) and $\nu=1/5$ (bottom panel).
            The potential energy per particle, $\langle \hat{V} \rangle/N$ 
            is plotted as a function of $1/\sqrt{N}$
            for systems with 
    %   $N=16, 36, 64, 100, 144$ and $196$ electrons.
        $N=36, 64, 100, 144, 196$ and $400$ electrons.
	Full circles:  correlation energies calculated by the standard
	method described above, the full line is a 
        least-square fit
  %	simple fit 
        [Eq.\ (\protect\ref{fit_std3}) and \ (\protect\ref{fit_std5})]
        used to extrapolate to the
	thermodynamic limit.
            Energies are in units of $e^2/l_0$. }
\label{fit3fit5}
\end{figure}

To get the the bulk (thermodynamic estimate) of the correlation
energy per particle one needs to perform a careful extrapolation of the results
as illustrated in  
Fig.\ \ref{fit3fit5}
where we show 
the correlation energy per particle for states $\nu=1/3$ and $1/5$ plotted
as a function of $1/\sqrt{N}$ for systems with 
$N=36, 64, 100, 144, 196$ and $400$ electrons.

We fitted the energies of Table~\ref{tabstandard}
for 
    $N=4, 16, 36, 64, 100, 144, 196$ and $400$ electrons
to a polynomial function as reported  in Ref.\ \cite{morf}
and obtained:
\begin{equation}
\frac{\langle \hat{V} \rangle_{1/3}}{N}= 
    \left( -0.4094+ \frac{0.0524}{\sqrt{N}}-\frac{0.0225}{N}  \right)
    \frac{e^2}{l_0}  \,,
\label{fit_std3}
\end{equation}
\begin{equation}
\frac{\langle \hat{V} \rangle_{1/5}}{N}=
       \left( -0.3273 + \frac{0.0200}{\sqrt{N}} - \frac{0.0172}{N} \right)
        \frac{e^2}{l_0}  \ .
\label{fit_std5}
\end{equation}
These interpolation lines are used to estimate the correlation energy
per particle
in the thermodynamic limit (the first term in each of the parentheses). 
Our results for thermodynamic limit are similar to those 
found in Ref.\cite{levesque},
$-0.4100 \pm 0.0001$ and
$-0.3277 \pm 0.0002$  
(in units of $e^2/l_0$) derived with the use of
the pair correlation function evaluated from MC simulations
with up to $N=256$ electrons and
generating as many as 5 million MC configuratons. 
Note how slow the convergence is,
although the extrapolation to $N \rightarrow \infty$ seems unambiguous
it is still time consuming and cumbersome
( even for $N=400$ the error is still circa $\sim 1$ \% ).

% ----------- rho ---------------------
\subsection{The one-particle density}      

Other physical quantities of interest that may be readily computed are
the single-particle density function and the pair distribution
function.
Given that the Laughlin wave function describes an isotropic liquid and
is rotationally invariant, the single-particle density depends only
on the radial distance from the center of the disk.  We may compute
the single-particle density by counting the number of 
electrons $N_{l}(\Delta r)$
found in several 2D shells of width $\Delta r$ centered around a
discrete set of distances  to the center $r_l = (l + \frac{1}{2})
\Delta r$  ($l = 0, 1, \ldots$):                           
\begin{equation}
\rho(r_l) \equiv
\left\langle
\frac{N_l(\Delta r)}{\Omega_l(\Delta r)} \right\rangle
\ ,
\label{single}
\end{equation}
where $\Omega_l(\Delta r) = \pi (\Delta r)^2 [(l+1)^2-l^2]$ is the area
of each 2D shell.  In the $\Delta r \rightarrow 0$ the quantity computed
corresponds unequivocally to the electron density
\begin{equation}
\rho(r) = \left\langle \sum_{i=1}^{N} \delta(r-r_i) \right\rangle \,.
\label{singledelta}
\end{equation}
 
\begin{figure}[!htb]
\includegraphics{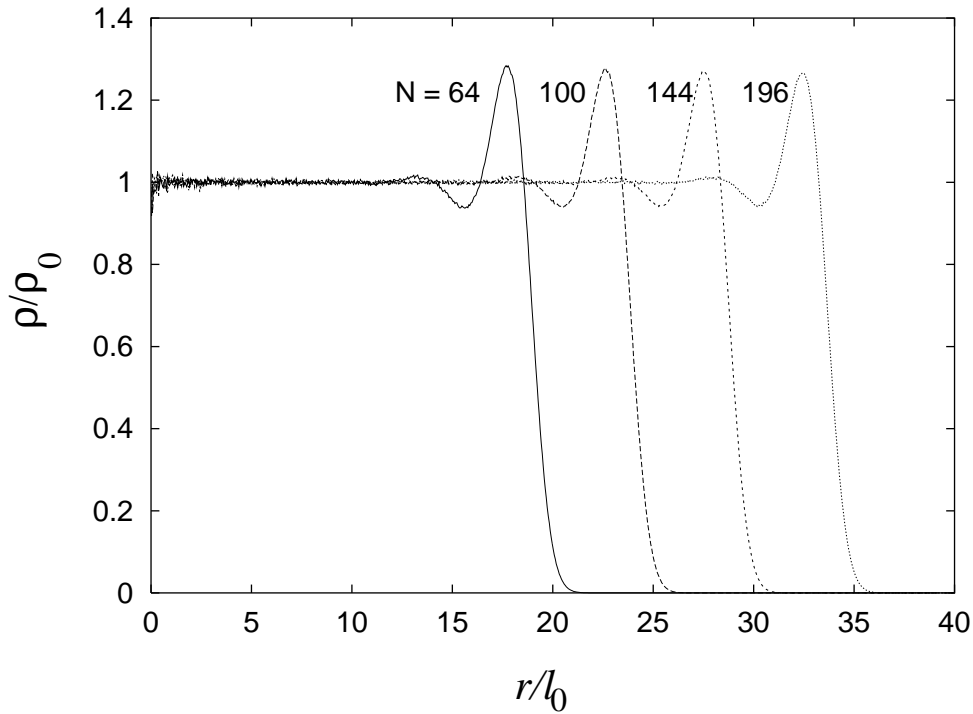}
\caption[]{One-body density function, $\rho(r)/\rho_0$, in the
           Laughlin state $\nu=1/3$ as a function of the
           distance $r/l_0$ from the center of the disk for
           systems with $N=64, 100, 144$ and $196$ electrons. Note the
persistence of an ``edge region'' of finite width and the development
of a ``bulk region'' for large $N$.  A discretization interval
$\Delta r = 0.05 \, l_0$ was used. }
\label{figrho}
\end{figure}

The computation of the single-particle density in the Laughlin state,
indicates a significant nonuniformity near the boundary (see Fig.\
\ref{figrho}).  As the number of electrons increase, a significant
portion of the system becomes uniform as expected.  Note, however,
that the non-uniformity near the edge always persists.  This behavior
can be used to characterize which electrons are ``in the bulk.''

%---------------------------- g(r) -----------------------------
\subsection{The pair distribution function}             

Another important quantity related with the trial wave function is
the pair distribution function, which corresponds to the conditional
probability density to find an electron at a distance $r$ from another
electron. For any homogeneous and isotropic liquid with uniform
density $\rho_{0}$ it is defined as
\begin{equation}
\rho_0 \ g(r)=\frac{1}{N} \left\langle \sum_{i=1}^{N} \sum_{j \neq i}^{N}
\delta (r-|\rr_i-\rr_j|) \right\rangle  \ .
\label{pair}
\end{equation}
Following the same procedure as above, we discretize in concentric
shells around the $i$-th electron and count the number of electrons
$N_{l}(\Delta r)$ in each shell, which should give $g(r)$ as $\Delta r
\rightarrow 0$ according to the following equation:
\begin{equation}
\rho_0 \ g(r_l)=\frac{1}{N} \frac{1}{\Omega_l(\Delta r)}
 \left\langle \sum_{i=1}^{N} \sum_{j \neq i}^{N} N_{l}(\Delta r)
 \right\rangle  \ .
\label{pair2}
\end{equation}
It is evident that electrons near the edges of the system could
contribute expuriously to these sums as their ``surroundings'' are
considerably different than those at the bulk.  To eliminate as much
as possible any boundary effects, it is convenient \cite{morf} to
consider only (for the ``$i$-electrons'' above) the electrons that are
within a small circle of radius $R_1$ around the origin.
If $N_1$ is the average number of electrons that are within this
small circle, then the approximation $\hat{g}(r_l)$ for the pair
distribution $g(r_l)$ is:
\begin{equation}
\rho_0 \ \hat{g}(r_l)=\frac{1}{N_1} \frac{1}{\Omega_l(\Delta r)}
 \left\langle \sum_{i=1}^{N_1} \sum_{j \neq i}^{N} N_{l}(\Delta r)
 \right\rangle  \ ,
\label{pairap}
\end{equation}                                      
where in this expression one is considering the pairs between any
electron $i$ ($i=1,\ldots,N_1$) lying inside the circle of radius $R_1$,
with all other electrons $j$ ($j = 1, \ldots,i-1,i+1,\ldots,N$) that may
lie either inside, or outside that circle.  This guarantees that 
the evaluation of $\hat{g}(r_l)$ involves only pairs, where at least
one member lies inside a circle of radius $R_1$ around the origin,
where correlations are believed to be close to those in the bulk of an
infinite system.

\begin{figure}[!htb]
\includegraphics{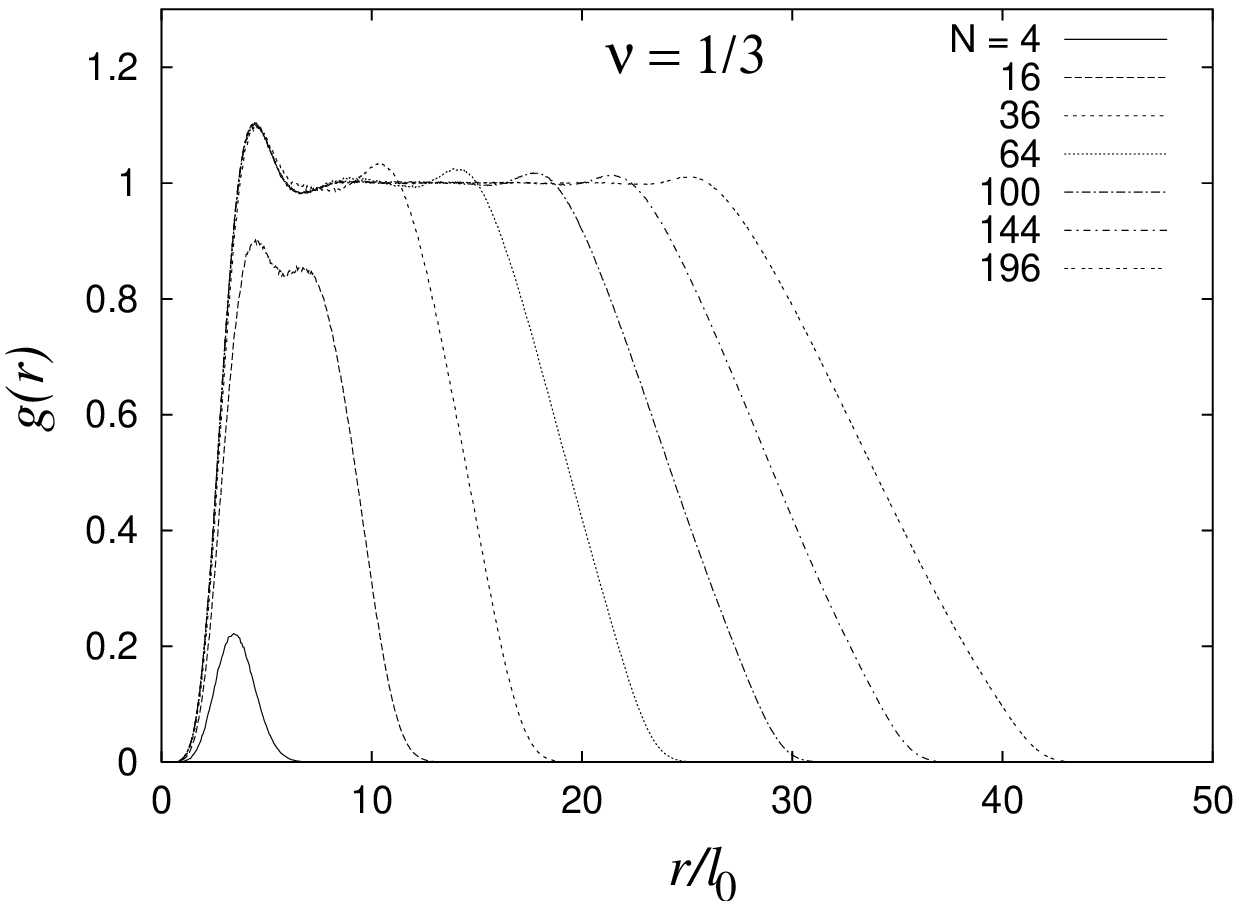}
\includegraphics{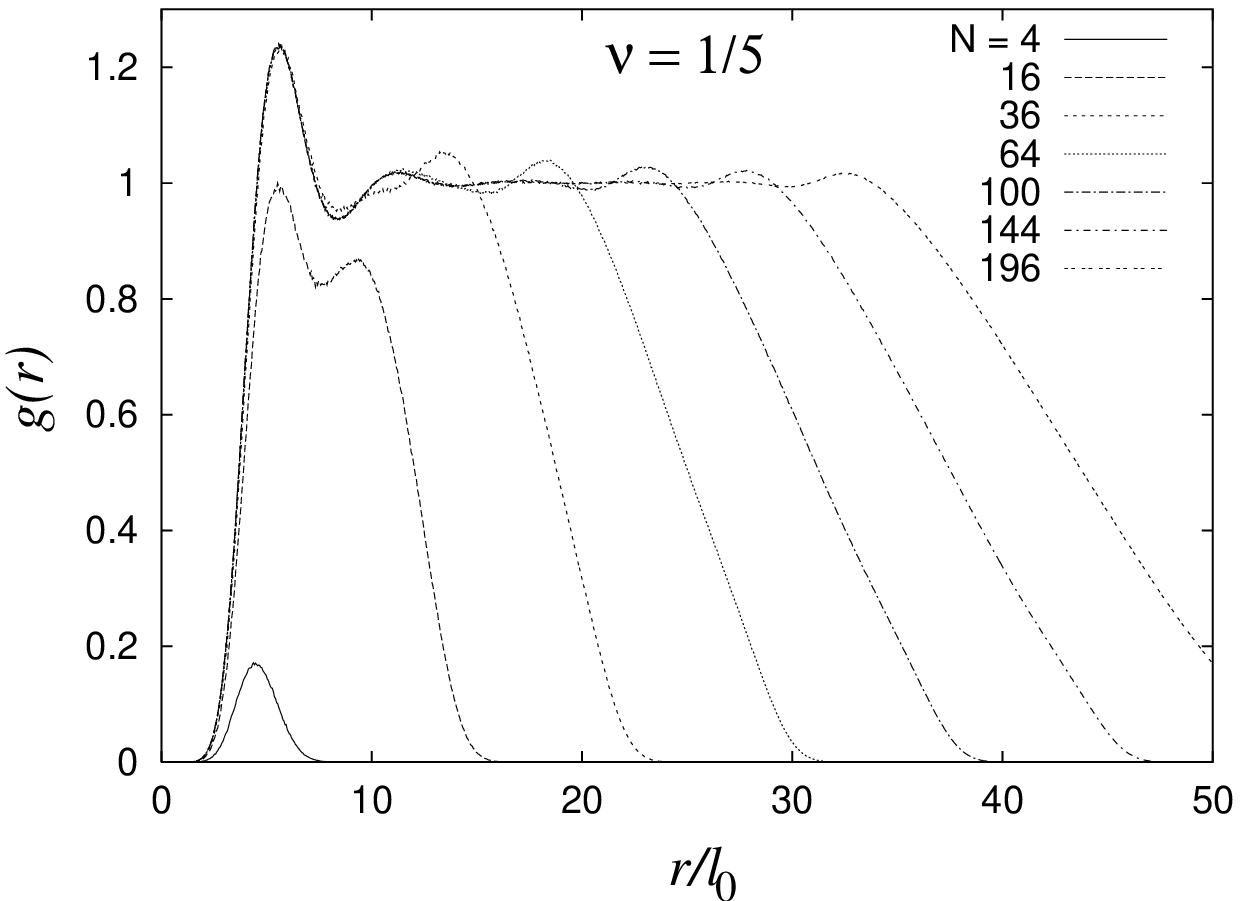}
\caption[]{Pair distribution function for $\nu=1/3$ 
            and $1/5$
            obtained by a standard Monte Carlo simulation in disk geometry
            for systems of $N=4, 16, 36, 64, 100, 144$ and $196$
            electrons.}                                              
%\label{g3standard}
\label{g35standard}
\end{figure}

Fig.\ \ref{g35standard} 
show plots of the pair distribution function for the 
states $\nu=1/3$ and  $1/5$ for
systems with $N=4, 16, 36, 64, 100, 144$ and $196$ electrons.
For our MC simulations we chose $R_1 = 0.25 \, R_{N}$ and a discretization
interval $\Delta r=0.05 \, l_0$. Note the gradual decay at large $r$
which reflects the finite size of the system.
 
The determination of the pair correlation function for a given
finite $N$ tends to be quite time-consuming but provides 
for an alternative way to compute the correlation energy per particle 
in the thermodynamic limit \cite{morf} by using
the formula:
\begin{equation}
\frac{\langle \hat{V} \rangle}{N}=\frac{\rho_0}{2}
      \int_{0}^{+\infty} d^2r \ \frac{e^2}{r} \left[g(r)-1\right]  \ ,
\label{thermog}
\end{equation}
which is valid in the limit of an infinite system.
Although the pair distribution function is obtained from a system
with a finite number of particles, one can calculate the 
thermodynamic value of the correlation energy per particle to a very 
good accuracy by using the slightly modified formula
\begin{equation}
\frac{\langle \hat{V} \rangle}{N}=\frac{\rho_0}{2}
      \int_{0}^{R_{cut}} d^2r \ \frac{e^2}{r} \left[\hat{g}(r)-1\right]  \ ,
\label{thermocut}
\end{equation}
in connjuction with the normalization condition
\begin{equation}
\rho_0 \ \int_{0}^{R_{cut}} d^2r \ \left[\hat{g}(r)-1\right]=-1  \ ,
\label{gnorma}
\end{equation}
which defines an upper cuttoff $R_{cut}$.  This approach produces
good estimates for the thermodynamic correlation energy per particle
as long as $\hat{g}(r)$ is able to reach its asymptotic value [$\hat{g}(r)
\simeq 1$].  Reasonable results can be achieved even for systems
of $N \geq 36$ electrons.

%%%%%%%%%%%%%%%%%%%%%%%%%%%%%%%%%%%%%%%%%%%%%%%%%%%%%%%%%%%%%%%%%
\section{An alternative method}
\label{sec:center}    

In the standard MC approach one needs to calculate
the expectation values of various quantities several times
for different $N$ in order to extract the 
thermodynamic estimate 
from the data by performing a $1/\sqrt{N}$ fit (and  taking
the limit $1/\sqrt{N} \rightarrow 0$ , as shown, e.g.\ in Fig.\
 \ref{fit3fit5}.
It is highly desirable to obtain estimates for the
bulk regime (in thermodynamic limit) without needing to perform
the above analysis.
In the following we describe a method that allows us to obtain
results consistent with the bulk regime, by doing simulations with
only a finite (relatively small) number of particles.
The method although approximate, yields very accurate
results corresponding to the bulk regime even when simulations
are performed for a small number of electrons.
The estimates are very stable over a wide range of $N$ and the
technical application of the simulation is less involved.

In order to obtain reliable estimates for the bulk regime,
we need to exclude from consideration the boundary-affected
outer region of the finite disk.
In a standard MC simulation the electrons are distributed
all over the 2D space, at a given instant it is obvious that
the electrons close to the central region of the disk resemble to
the bulk regime better than those close to the boundary.
However during the simulation each of the previously
``bulk'' electrons moves around the whole disk therefore
the correlation energy of such electron with the other electrons
is not a good estimate of the correlation energy in the bulk regime.

The core of the method proposed here is to consider an electron pinned 
to the center of the disk which, by construction is the point which
most closely resembles the bulk of the system.  Therefore, if we are
able to derive results where only the correlation energy of that
particular electron with the rest (away from edge) is involved, 
we anticipate that such
estimates should approximate the bulk regime much more accurately than 
other methods, and as a result thermodynamic limit values can be achieved
even in a system with a relatively small number of electrons.
%-----------------------------------------------------
One has to recall that for a finite system, the Laughlin wave 
function describes an
incompressible system of strongly correlated ellectrons with
uniform density $\rho_0$ only in the bulk region 
(central part of the disk not very close to the boundary), while
close to the boundary (where the density of electrons falls to zero)
the fluid becomes compressible and there is a deviation of the
electron density from its constant value in the bulk. 
%
%---------------------------------------------------------------------

In our MC simulations, we consider a Laughlin-like state [Eq.\
(\ref{laughlin})] in which one electron is pinned at the position
$z_0$ (we consider $z_0=0$), and $N' = N-1$ electrons are free
to move (i.e.\ in a typical MC step):
\begin{equation}
\Psi^{\prime}_{m}(z_0,z_1, \ldots z_{N^{\prime}})=
\Psi_{m}(z_1, \ldots z_{N^{\prime}}) \ 
\prod_{j=1}^{N^{\prime}} (z_j-z_0)^m \ e^{-\frac{|z_0|^2}{4 l_0^2}} \ .
\label{laughlinsimp}
\end{equation}
When an attempt is made to move electron $i$ ($i = 1, \ldots, N'$)
from $\rr_{i}^{\rm old}$ to $\rr_{i}^{\rm new}$, the MC
probability ratio is given by
\begin{equation}
\frac{|\Psi^{\prime}(\rr_0, \ldots \rr_{i}^{\rm new} \ldots
\rr_{N^{\prime}})|^2}{
      |\Psi^{\prime}(\rr_0, \ldots \rr_{i}^{\rm old} \ldots
\rr_{N^{\prime}})|^2}= 
\frac{|\Psi(\rr_1, \ldots \rr_{i}^{\rm new} \ldots \rr_{N^{\prime}})|^2}{
      |\Psi(\rr_1, \ldots \rr_{i}^{\rm old} \ldots
\rr_{N^{\prime}})|^2} \ e^{ m \left(\ln |{\rr_i}^{\rm new}-\rr_{0}|^2 
                    -\ln |{\rr_i}^{\rm old}-\rr_{0}|^2 
     \right) } \ .
% \ \ \ ; \ \ \ \rr_0=0 \ .
\label{probsimple}
\end{equation}

Since the electrons are identical we need only consider the average
correlation energy of one specified electron with the rest of the electrons
to compute the electron-electron interaction energy, therefore to
this level of simplification the electron-electron energy per
particle is obtained by considering {\em only} the interaction between
the pinned electron and the rest of $N_i$ other electrons contained
within a inner disk of radius $R_i < R_N$ where the electron density
is approximately equal to the bulk value $\rho_0$.
We found that a reasonable choice for the radius of inner disk that
excludes the edge electrons is $R_i=0.75 \, R_N$, therefore this
value was adopted in all the following simulations.

To be consistent with the above procedure also the disk's
electron-background and background-background
energy should be calculated  within the same degree of simplification 
(see Appendix %\ref{appendix}
).
It is, therefore,
useful to first calculate the one-body density function (see e.g.\
Fig.\ \ref{figrho}) in order to determine an optimal $R_i$ for
future use.
If $R_i$ is reasonably large, we expect the correlation energy
calculated in this way to closely correspond
to the desired correlation energy per particle in 
the thermodynamic limit (see Appendix):
\begin{equation}
\frac{\langle \hat{V} \rangle}{N} \simeq
\frac{1}{2} \left\langle \sum_{i=1}^{N_i} 
        \frac{e^2}{|\vec{r}_i-\vec{r}_0|} \right\rangle-
        \sqrt{\frac{N_i+1}{2 m}} \ \frac{e^2}{l_0}  \ .
\label{correl_z0}
\end{equation}
As in the standard MC method case, our MC runs consist of
100,000 discarded equalibration MCS-s followed by
$2 \times 10^6$ MCS-s for averaging purposes.                      

In Table \ref{tabcenter} we show the correlation energy per particle
for finite systems of $N$ electrons and Laughlin states $m=3$ and
$m=5$ calculated by pinning an electron at $z_0 = 0$ as described
above [see Eq.\ (\ref{probsimple})], and using Eq.\ (\ref{correl_z0}).
The results are rounded in the last digit.

\begin{figure}[!htb] 
\includegraphics{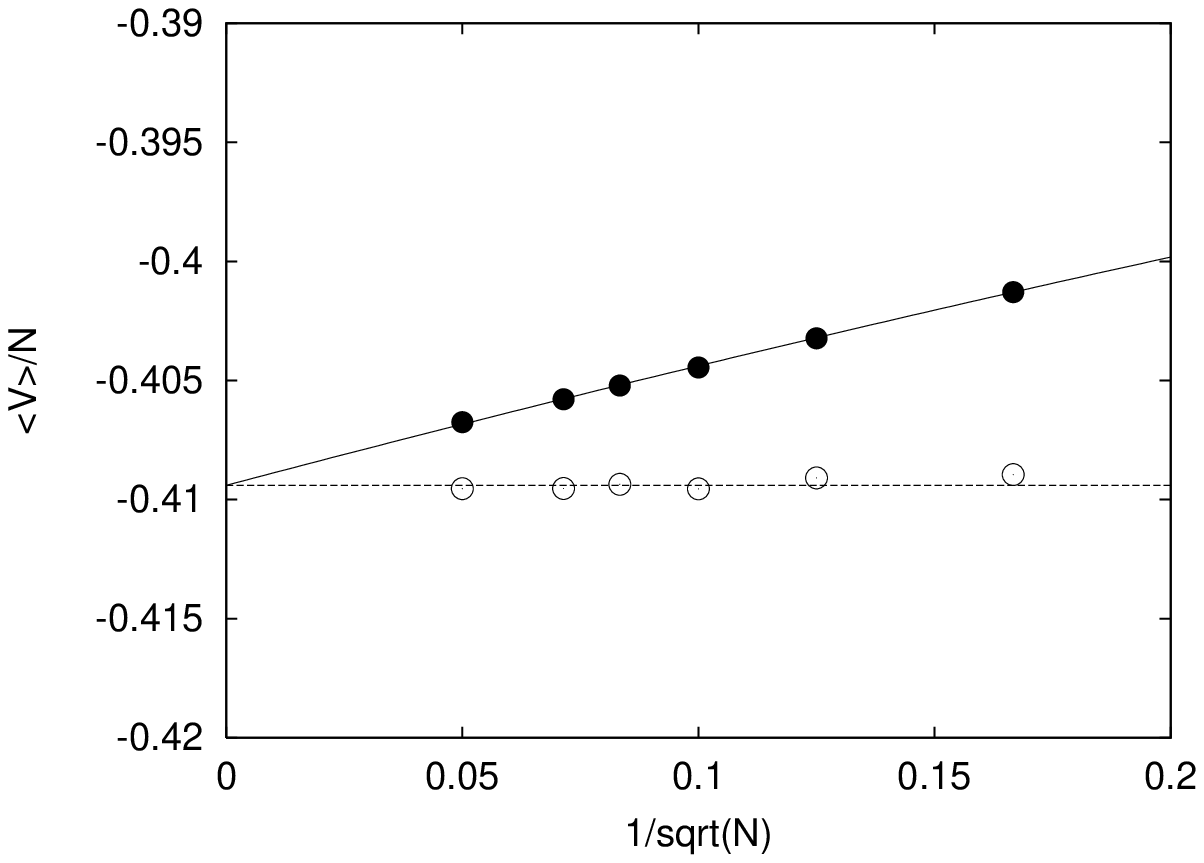}
\includegraphics{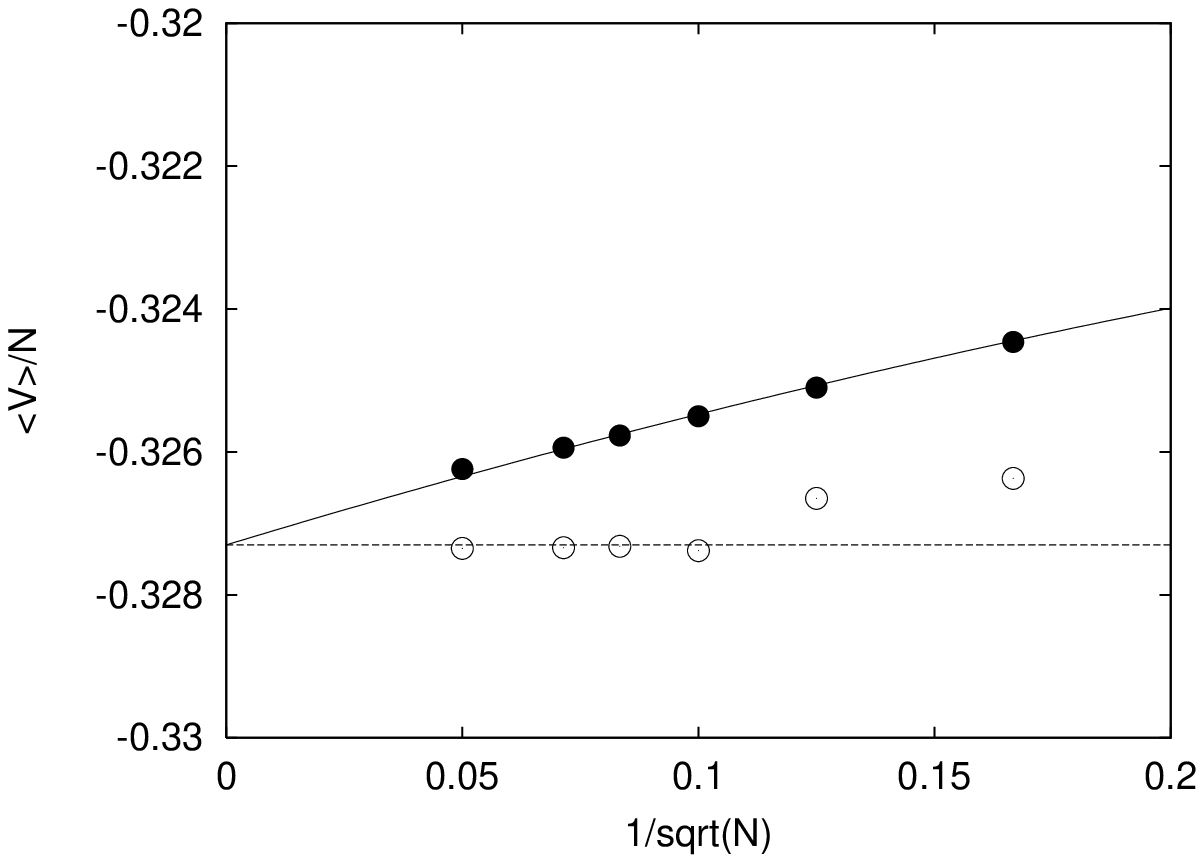}
\caption[]{ Monte Carlo results in disk geometry for the
            Laughlin state at 
            $\nu=1/3$ (top panel)and $1/5$ (bottom panel). 
            The potential energy per particle, $\langle \hat{V} \rangle/N$ 
            is plotted as a function of $1/\sqrt{N}$
            for systems with 
      $N=36, 64, 100, 144, 196$ and $400$ electrons.
	Full circles:  correlation energies calculated by the standard
	method (Sec.\ \protect\ref{sec:standard})
        the full line is a least-square fit
        [Eq.\ (\protect\ref{fit_std3}) and (\ref{fit_std5})] 
        used to extrapolate to the
	thermodynamic limit.
	Empty circles: correlation energies calculated by the alternative
	method described in Sec.\ \protect\ref{sec:center},
       the dashed line is a visual aid indicating the
       thermodynamic limit.  
        Note how
	the thermodynamic limit is approached faster in the second method.
            Energies are in units of $e^2/l_0$. }
\label{compare35}
\end{figure}

\begin{table}
\caption[]{Correlation energy per particle in the Laughlin state
           for filling factors $\nu=1/3$ and $1/5$, obtained via
           a Monte Carlo simulation in a disk geometry using the method
           of pinning one electron at the center of the disk.
           Energies are in units of $e^2/l_0$. }
\label{tabcenter}
\begin{center}
\begin{tabular}{|c|c|c|}
\hline
N         & m=3                & m=5                          \\ \hline
4         & -0.38187      & -0.30157     \\ \hline
16        & -0.40898      & -0.32722      \\ \hline
36        & -0.40895      & -0.32637      \\ \hline
64        & -0.40909      & -0.32665      \\ \hline
100       & -0.40955      & -0.32738      \\ \hline
144       & -0.40936      & -0.32732     \\ \hline
196       & -0.40953      & -0.32734      \\ \hline
%--------------------
%400       & -0.40954      & -0.32739      \\ \hline
400       & -0.40954      & -0.32735      \\ \hline
\end{tabular}
\end{center}
\end{table}
%
%
%Figs. \ref{fit3} and  \ref{fit5} 
In Fig. \ref{compare35} we show the potential energy per particle 
for $\nu=1/3$ and $1/5$
computed from the alternative method and plotted as a function
of $1/\sqrt{N}$.
For the sake of comparison we also plot the results from the
standard method (Sec. \ref{sec:standard}).

%-----------------------------------------------------------------
%

It is striking to note how much faster the alternative method
converges to the thermodynamic limit.
Differently from the standard MC approach, the alternative method
that we introduced does not need to have the data points
least-square fitted to get the thermodynamic limit value.
One merely needs to choose a big enough $N$ (for instance $N=196$)
and do a MC run that typically will generate bulk (thermodynamic limit) 
results to an excellent degree of accuracy.

%----------Pair distribution function
\subsection{The pair distribution function}     

\begin{figure}[!htb]
\includegraphics{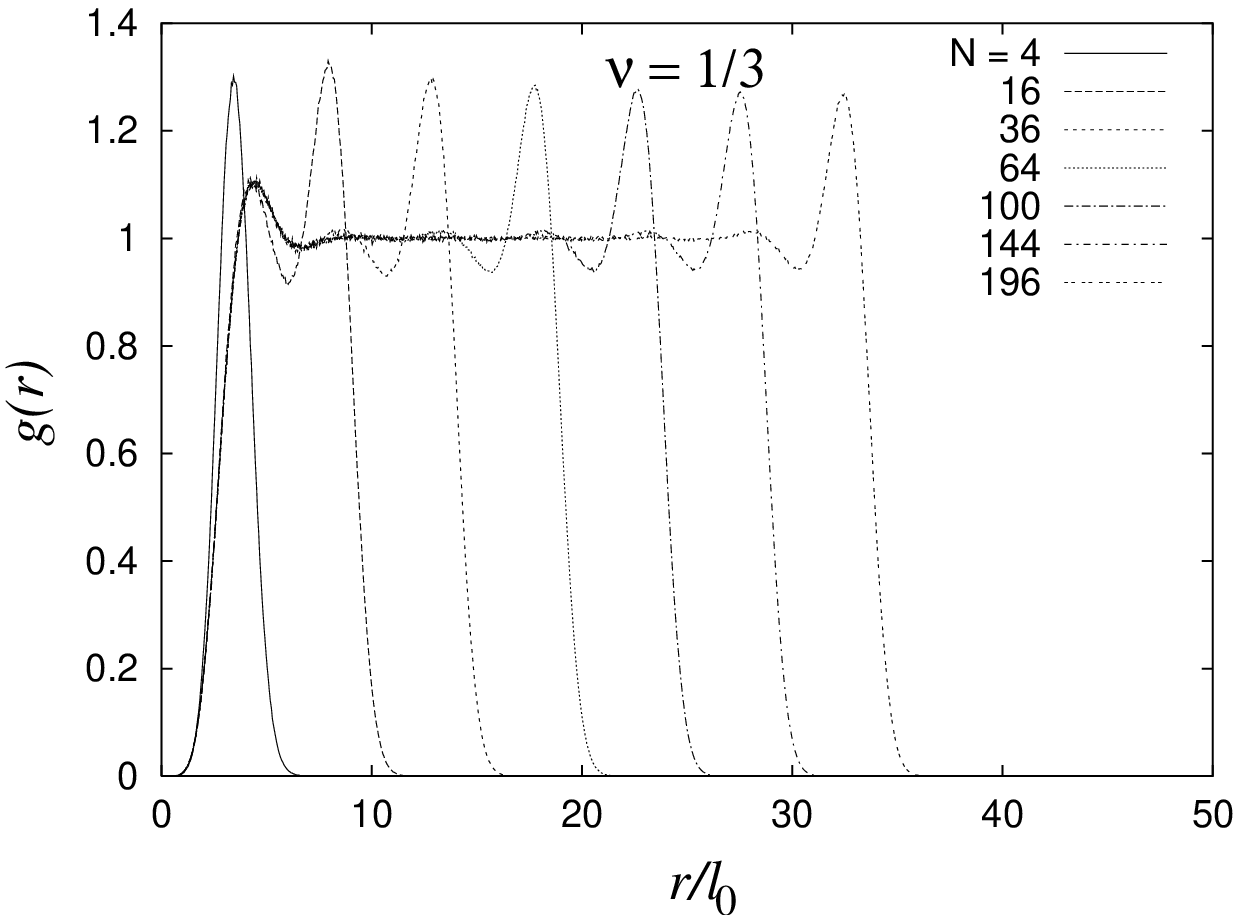}
\includegraphics{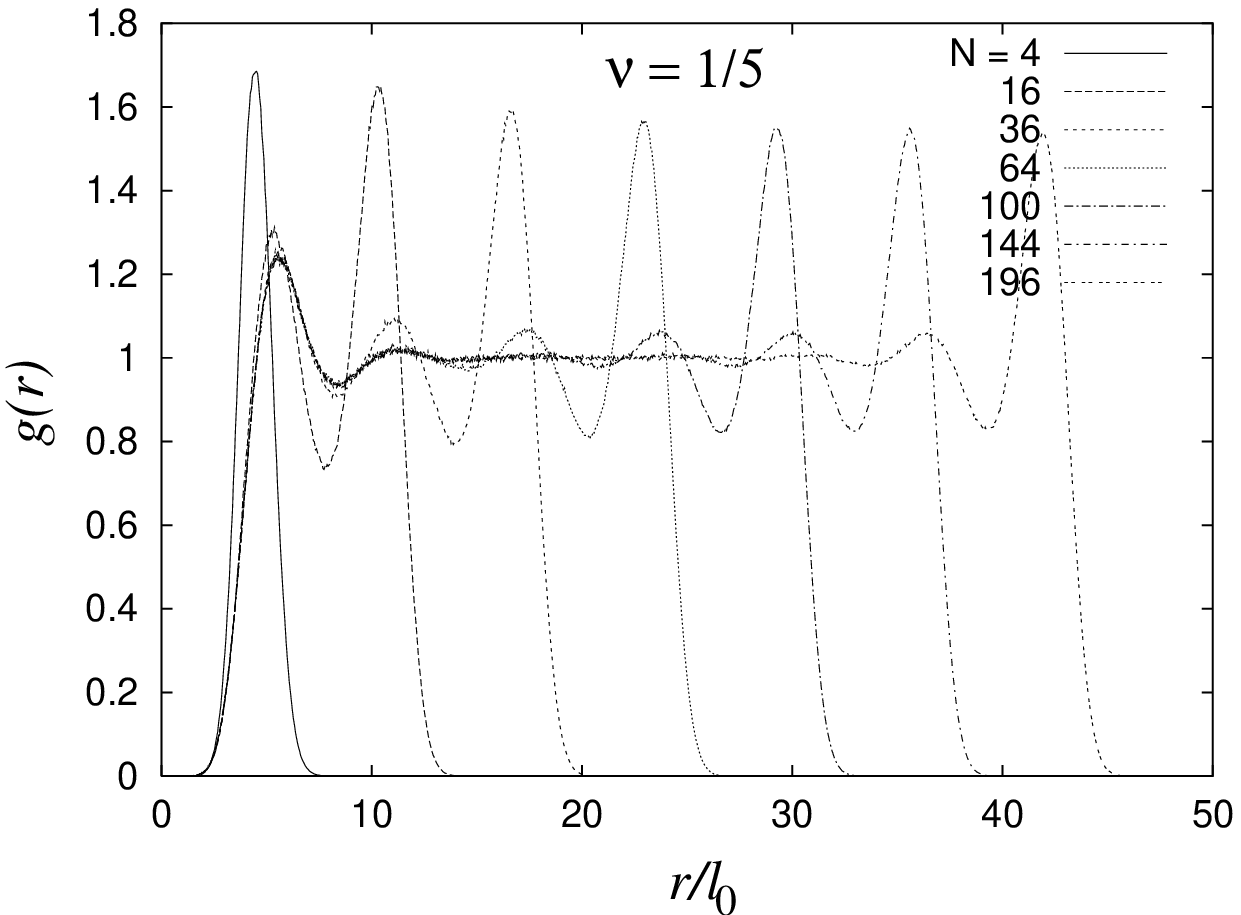}
\caption[]{ Pair distribution function for the state $\nu=1/3$ 
            and $1/5$
            obtained after a Monte Carlo simulation in the disk geometry
            for systems of $N=4, 16, 36, 64, 100, 144$ and $196 $
            electrons 
            with one electron held fixed at the center of the disk . }
\label{g35center}
\end{figure}

Using the same ideas presented above (keep one electron pinned at
$z_0=0$) the pair distributiuon function is very easily calculated.
The essence of the method consists in measuring the one-particle
density {\em excluding} the pinned electron which is obviously the
pair distribution function instead of considering all the possible
pairings between the electrons.  As before, the advantage of this
method, besides its simplicity, resides in the fact that this
electron, being the farthest from the edge, is in an environment
closest to that in a bulk system.  Therefore, one has to calculate:
\begin{equation}
\rho_0 \ g(r) \simeq
\left\langle \frac{N_l(\Delta r)}{\Omega_l(\Delta r)} \right\rangle \ ,
\label{pairsimple}
\end{equation}                         
where $N_l(\Delta r)$ is the number of electrons found in the 2D shell
$\Omega_l(\Delta r)$ with distance range $(r_l,r_l+\Delta r)$ from the
pinned electron at the center of the disk.
In Fig.\ \ref{g35center}
we show
plots of the pair distribution function for the states $\nu=1/3$ and
$\nu=1/5$ for systems with $N=4, 16, 36, 64, 100, 144$ and $196$
electrons obtained with a choice of $R_i=0.75 \, R_N$
and using a discretization interval $\Delta r = 0.05 \, l_0$.

%%%%%%%%%%%%%%%%%%%%%%%%%%%%%%%%%%%%%%%%%%%%%%%%%%%%%%%%%%%%%%%%%%%%%%
\section{Summary and discussions}
\label{sec:discussion}

We implemented an alternative Monte Carlo method to calculate the 
properties of Laughlin states of the fractional quantum  Hall effect
in the thermodynamic limit while using a very small number of electrons.
The key point of this method is the pinning of an electron in the
center of the disk, so that the potential energy and correlation
functions calculated through the pinned electron accurately represent 
the bulk (thermodynamic limit) even in a relatively small system.
The idea is quite general and, in principle, can be applied to any
system as far as the main concern is the calculation of
correlation effects such as the potential energy, the pair distribution
function, etc.
For systems such as 2D electronic one-component-plasmas~\cite{caillol},
composite fermion states described by the Jain's 
unprojected wave function~\cite{jain} etc, that is all that matters.
Obviously such alternative method can always be used
to calculate the potential energy and related quantities of other
more diverse systems, with the the kinetic energy calculated in the 
standard way whenever applicable.
By using this alternative method we analyzed the properties of the Laughlin 
states corresponding to filling factor $\nu=1/3$ and $1/5$ by
performing Monte Carlo simulations in disk geometry for systems
with up to   $N=400$ electrons.
The correlation energy per particle and the pair distribution function 
computed in
this approach are compared to corresponding bulk-regime values
obtained via a standard Monte Carlo simulation in disk geometry, where
a careful extrapolation in thermodynamic limit has been performed.
We find that such approach allows us to obtain accurate
bulk regime (thermodynamic limit) values for various quantities
using a modest number of electrons 
( even for $N=16$ the error is less than $0.1 \ \%$ ).

\section{Acknowledgments}

We would like to thank A.T.\ Dorsey and M.\ Fogler for 
useful discussions.
Acknowledgement is made to the University of Missouri Research Board
and to the Donors of the Petroleum Research Fund, administered by the
American Chemical Society, for support of this research.

% -------------------------------------------------------------------------
\section{Appendix}
%\section*{}
%\label{appendix}                                 

Let us consider a system of $N$ interacting 2D electrons coupled
to the positive neutralizing background that fills a finite disk
and guarantees overall charge neutrality.
The expectation value of the electron-background potential 
energy per particle can be written as:
\begin{equation}
\frac{\langle \hat{V}_{eb} \rangle}{N}=-\frac{\rho_0}{N} 
\int d^2r_1 \rho(\rr_1) \int_{\Omega_N} d^2r \
\frac{e^2}{|\rr_1-\rr|}  \ ,
\label{ebintegral}
\end{equation}
where  $\rho(\rr_1)$ is the one-body (single) electron density
function given by
\begin{equation}
\rho(\rr_1)=N
\frac{\int d^2r_2 \ldots d^2r_N \ |\Psi(\rr_1 \ldots \rr_N)|^2}
     {\int d^2r_1 \ldots d^2r_N \ |\Psi(\rr_1 \ldots \rr_N)|^2}
\ \ \ ; \ \ \
\int d^2r_1 \  \rho(\rr_1)=N  \ .
\label{onebody}
\end{equation}
One notes that when the one-body electron density becomes uniform,
$\rho(\rr_1) \approx \rho_0$ and the system is sufficiently
large so that most of the electrons are to be found inside
the finite disk 
then
$-\langle \hat{V}_{eb} \rangle/{N} \approx 
2  \langle \hat{V}_{bb} \rangle/{N} $, 
therefore we would have:

\begin{equation}
\frac{\langle \hat{V}_{eb} \rangle+\langle \hat{V}_{bb} \rangle}{N} 
\simeq
\frac{1}{2} \frac{\langle \hat{V}_{eb} \rangle}{N} \ .
\label{background}
\end{equation}
By making a
preliminary calculation of the one-body density function 
we could estimate the radius, $R_i < R_N$ of an inner disk 
where the electrons have supposedly uniform density 
therefore we could argue that all these electrons inside 
this inner disk are in the bulk regime.
If $N_i$ is the number of electrons within this reference circle
of radius $R_i$ in addition to the electron pinned at the
center of the disk then a 
 good estimate for the potential energy per particle in the
thermodynamic limit is obtained from the quantity:
\begin{equation}
\frac{\langle \hat{V} \rangle}{N}=\frac{1}{N_i+1}
\left\langle  \sum_{i=0}^{N_i} \sum_{i<j}^{N_i}
 \frac{e^2}{|\rr_i-\rr_j|}
\right\rangle+
\frac{1}{N_i+1} \left\langle \sum_{i=0}^{N_i} \hat{v}_{eb}(\rr_i)
              \right\rangle+
\frac{\langle \hat{V}_{bb} \rangle}{N_i+1} \ ,
\label{eepoti2}
\end{equation}
where $\langle \hat{V}_{bb} \rangle/(N_i+1)$ is the 
background-background energy per particle of a positive charge that exactly
neutralizes the charge of $N_i+1$ electrons.
Since the electron-electron potential energy per particle was obtained 
to a level of simplification where only the interaction between the pinned
electron and other $N_i$ electrons was considered, then
the whole 
potential energy per particle should
be calculated in the same grounds too, namely, using as reference
only the interaction energy of the pinned electron with the positive
background:
\begin{equation}
\frac{\langle \hat{V} \rangle}{N}=
\frac{\langle \hat{V}_{ee} \rangle}{N}+
\frac{\langle \hat{V}_{eb} \rangle+\langle \hat{V}_{bb} \rangle}{N} 
\simeq
\frac{1}{2} \left\langle \sum_{i=1}^{N_i} 
        \frac{e^2}{|\vec{r}_i-\vec{r}_0|} \right\rangle+
\frac{1}{2} \left \langle \hat{v}_{eb}(\rr_0) \right \rangle =
\frac{1}{2} \left\langle \sum_{i=1}^{N_i} 
        \frac{e^2}{|\vec{r}_i-\vec{r}_0|} \right\rangle
-\frac{\rho_0}{2} \int_{\Omega_{N_i}} d^2r \ \frac{e^2}{r} \ .
\label{ebpinned}
\end{equation}

As a result
the correlation energy per particle 
in the thermodynamic limit can be written as in Eq. (\ref{correl_z0}).
Note that use of relation
$-\langle \hat{V}_{eb} \rangle/{N} \approx 
2  \langle \hat{V}_{bb} \rangle/{N} $ to express
$(\langle \hat{V}_{eb} \rangle+\langle \hat{V}_{bb} \rangle)/{N} $
in Eq.(\ref{background})
in terms of $\langle \hat{V}_{bb} \rangle/N$ is inaccurate in view of the
approach adopted in our method therefore should be avoided.

%--------------------------------------------------------------------

\end{document}